\documentclass[conference]{IEEEtran}

\usepackage{array}

\ifCLASSINFOpdf
   \usepackage[pdftex]{graphicx}
\else
\fi

\begin{document}
%
\title{Sieving Fake News From Genuine: A Synopsis}



\author{\IEEEauthorblockN{Shahid Alam}
	\IEEEauthorblockA{\textit{Department of Computer Engineering} \\
		Adana Alparsalan Turkes Science and Technology \\ University, Adana, Turkey \\
		salam@atu.edu.tr}
\and
\IEEEauthorblockN{Abdulaziz Ravshanbekov}
	\IEEEauthorblockA{\textit{Department of Computer Engineering} \\
	Adana Alparsalan Turkes Science and Technology \\ University, Adana, Turkey \\
	ravshanbekov.abdulaziz@gmail.com}
}


%


\maketitle

\begin{abstract}
With the rise of social media, it has become easier to disseminate fake news faster and cheaper, compared to traditional news media, such as television and newspapers. Recently this phenomenon has attracted lot of public attention, because it is causing significant social and financial impacts on their lives and businesses. Fake news are responsible for creating false, deceptive, misleading, and suspicious information that can greatly effect the outcome of an event. This paper presents a synopsis that explains what are fake news with examples and also discusses some of the current machine learning techniques, specifically natural language processing (NLP) and deep learning, for automatically predicting and detecting fake news. Based on this synopsis, we recommend that there is a potential of using NLP and deep learning to improve automatic detection of fake news, but with the right set of data and features.
\end{abstract}


%
\IEEEpeerreviewmaketitle

\section{Introduction}

Fake news is not a new phenomenon, but recently it has attracted more public attention. The rise of social media, easier access, and faster and cheaper online dissemination of fake news compared to other traditional news media, such as newspapers and television, makes it particularly relevant in this new age of information.

Conroy et al. \cite{conroy2015} divides the approaches to detecting deception into two major categories, linguistic and network approach. In the linguistic approach researchers have used: analysis of n-grams of words (i.e., bag of words); syntax and semantic analysis of the text; and structural analysis of text to find incoherency between deceptive and truthful messages. In the network approach researchers have used: knowledge networks for fact-checking; and social network behavior, such as using the metadata and telltale behavior of questionable sources.

There are basically two major techniques manual and automatic detection of fake news. Manual detection is carried out by experts in the field or by crowds (a large number of regular people acting as fact checkers). Automatic detection relies on a combination of information retrieval and machine learning techniques. Manual detection is more reliable but not scalable. Automatic detection is less reliable but scalable. In this paper we only discuss automatic detection of fake news.

In the next Sections we explain what are fake news with examples and present and discuss the current machine learning techniques for automatically detecting fake news.

\section{Fake News}\label{fake-news}

Before the invention of paper in China, rumors (fake news) used to spread by word of mouth, and were mostly confined to a local community. The invention of paper made the dissemination of news easier. But still it used to take time for a news to reach/spread the far corners of a state/country. The emergence of Internet, social media and smart phones have revolutionized this process, and now almost everyone can claim that I have \emph{news in my pockets}. This phenomenon is very useful in making people more aware and knowledgeable of the events and their surroundings, but presents new challenges of \emph{sieving fake news from genuine} out of this large amount of information/data in our pockets. 

Fake news spread mostly through online media, in the form of text, images and videos. Fake news are false, and there purpose is to create deceptions and mislead people, so that it changes the outcome of an event. The events can be sending someone to jail, or manipulating the result of elections, etc. In this paper we only focus on fake news that are in the form of text. We formally define \emph{fake news} as follows. \\

\noindent DEFINITION: \emph{Fake news is any text data disseminated through online media, such as newspapers, websites, social networks, etc., that creates false, deceptive, misleading, and suspicious information which can significantly effect the outcome of an event.} \\

\noindent Some recent examples of fake news: \\

\begin{figure*}[htb]
	\centering
	\includegraphics[scale=0.5]{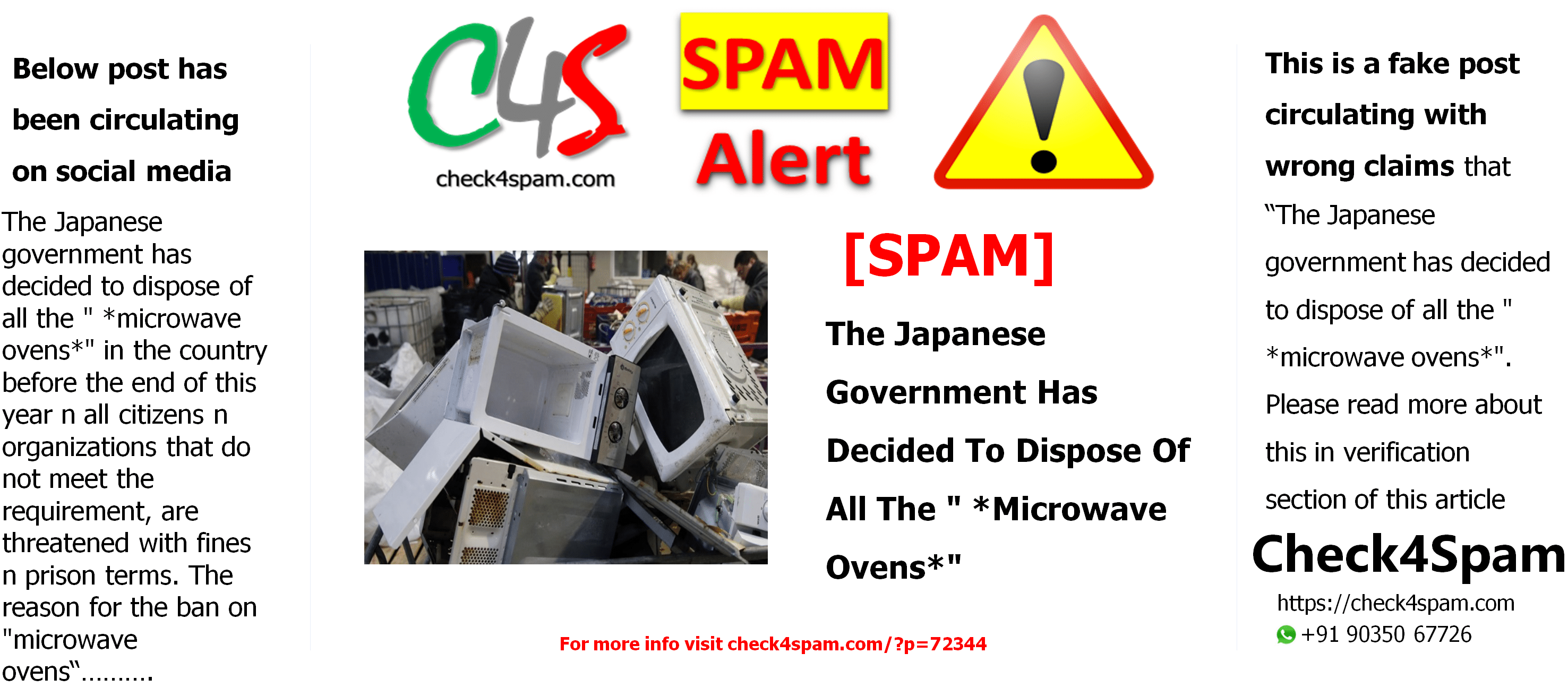}
	\caption{Example showing manual detection of a fake news about microwaves banned in Japan.}
	\label{fig:fake-news-Japan}
\end{figure*}

\noindent (1) The government of Japan announced that it was banning the use of microwave ovens in the country by 2020 \cite{FN-example-japan}: This fake news was originated on a Russian website about Japan abandoning the use of microwave ovens by 2020. The complete news was in Russian and when translated via Google it stated about banning microwave ovens and prisoning people not fulfilling the requirement. It caused panic in the mind of people. This news was verified to be fake by several fact checking web sites, such as by \emph{chek4spam.com} shown in Figure \ref{fig:fake-news-Japan}, as it was published by a satire website \emph{panorama.pub}. Fact checking web sites either use experts in the specific domain or use crowd sourcing for detecting fake news. This is an example, showing manual detection of fake news. \\

\begin{figure}[htb]
	\centering
	\includegraphics[scale=1]{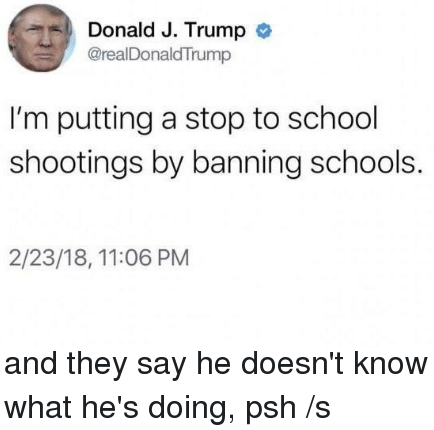}
	\caption{Example of a fake news about Donald Trump banning schools. \emph{psh} is an expression used when the writer disagrees with (urbandictionary.com).}
	\label{fig:fake-news-Trump}
\end{figure}

\noindent (2) President Trump to ban schools in order to stop school shootings \cite{FN-example-trump}: After a mass shooting at a school in Florida this fake news became viral on social media in February 2018. At the time of shooting, the U.S president Trump offered number of ideas, such as arming teachers, and raising the minimum age to purchase gun, but never suggested banning schools. This news was published on several satire websites and as a meme \footnote{Ideas, stories, phrases, etc., typically funny in nature, spread by Internet users, often with slight variations.} on \emph{me.me} \cite{FN-example-trump-original} as shown in Figure \ref{fig:fake-news-Trump}. \\

\noindent (3) Two altar boys put marijuana in the censer-burner of a Cathedral in Spain \cite{FN-example-censer}: This fake news went viral on social media in January 2018 in Spain. More than 2 million people read the news. As a consequence, they were detained overnight and released the other day without charge, but were fired from the Cathedral as altar boys. \\

\noindent (4) Barack Obama attended Columbia University as a foreign student \cite{FN-example-obama}: This fake news was created in 2012 U.S presidential elections about Barrack Obama, that he attended a college in U.S as a foreign student, to demonstrate that at some point in his life he was not a U.S citizen and is therefore ineligible to become the president of the United States. \\

From the above examples, we can see that these fake news have three things in common: (1) they became viral on social media; (2) they were deceptive and misleading; (3) they had the potential, and to some extent were able, to effect an event, such as causing panic and outcome of an election, etc.

\subsection{Natural Language Processing and Deep Learning}

\emph{Natural language processing} (NLP) is a branch of machine learning that deals with processing a natural (human) language, such as English, Spanish, French, and Chinese etc. NLP can be used for extracting important part of a text and providing an interpretation for that text. NLP is a hard problem, because unlike computer languages, natural languages are inherently ambiguous. There is a strong and natural relation between NLP and fake news (as defined in Section \ref{fake-news}) processing.

\emph{Deep learning} is one of the fields of machine learning. It involves learning from several layers. These layers consist of increasingly meaningful representations of input data. The depth of the model is the number of layers in the model. These layers of representations are learned through \emph{neural networks}. These networks map the input (such as an image) to the target (such as cat) during a deep sequence of simple data transformations (layers). A neural network is parametrized by its weights. A loss score is used as a feedback signal to adjust the weights. After several iterations the loss is minimized and the output is produced which is as close to the target as possible.

There are some advantages of applying deep learning to fake news detection such as:

\begin{itemize}
	\item
	The ability to learn feature representations rather than manually specifying and extracting features from the natural language, and can continually improve.
	
	\item
	Easily adapted to a new problem.
	
	\item
	Has shown great strength in processing text, speech and image, all of them are excessively used in fake news.
\end{itemize}

Besides some of the advantages listed above, to produce good results deep learning requires massive data and time for training.

Due to the natural relation between NLP and fake news, and strengths of deep learning listed above, recently \cite{yang2018early, volkova2017separating, NLP-rashkin2017truth, NLP-ruchansky2017csi, NLP-karimi2018multi, NLP-kirilin2018dsjm, NLP-pham2018study, NLP-della2018automatic, NLP-kochkina2018all, NLP-hanselowski2018ukp, NLP-bhattacharjee2017active, NLP-torabi2019big, NLP-borges2019combining} there is an interest in applying these two techniques for detecting fake news.

\section{State of the Art}\label{sec:related-work}

Here we discuss some of the previous recent works on predicting and detecting fake news using machine learning, specifically NLP and deep learning.

Kai et al. \cite{shu2019beyond} proposed a technique that exploits relationship among publishers, news pieces and users to predict fake news. For modeling this tri-relationship they presented a new framework \emph{TriFN} that employs a linear classifier. They assign each user a credibility score based on the user's online behavior. A user with a low credibility score is more likely to share fake news. Their classifier outperforms the baseline classifiers, such as Decision Tree, NaiveBayes, and Random Forest etc.

Yang et al. \cite{yang2018early} proposed a model for detection of fake news on social media through classifying news propagation paths. They capture the global and local variations of user characteristics to build a time series classifier. They compare their classifier with a series of baseline fake news detection classifiers, such as SVM, Decision Tree, and Random Forest etc. Their classifier combines both recurrent and convolutional neural networks in one, and therefore shows better results than the other compared classifiers in the paper \cite{yang2018early}.

Volkova et al. \cite{volkova2017separating} presented a technique that classify suspicious posts using linguistic and network features. They build a neural network model based on state of the art techniques, such as recurrent and convolutional neural networks. They also build a logistic regression model for comparison. They got better results with neural network model compared to logistic regression.

Siering et al. \cite{siering2016detecting} addressed detecting deception on crowdfunding platforms. They extract linguistic and content based features related to the different types of communication on these platforms. They examine both static and dynamic communications. Dynamic communication is carried out in real-time, i.e., it is analogous to face-to-face communication, and static communication is carried out in delayed mode. They also used different classifiers for testing their detection approach and achieve good results with Support Vector Machine (SVM) classifier.

Afroz et al. \cite{afroz2012detecting} used stylometry to detect deception in online writing. They selected a large number of (736) different features. The major type of features included lexical, syntactic, content specific, grammar and vocabulary complexity, uncertainty, author's attributes etc. For classification they used different classifiers and achieved good results with the SVM classifier. The results presented in the paper show that they were able to detect imitation attacks with 85\% accuracy and obfuscation attacks with 89.5\% accuracy.

Hannah et al. \cite{NLP-rashkin2017truth} used NLP to compare language of real news with fake news, such as hoaxes and propaganda, to find linguistic characteristics of untrustworthy text. A ratio was found for each linguistic feature, and refers to how frequently it appeared in the fake news compared to the real news. They build a Long Short-Term Memory (LSTM) model and compared it  with other Maximum Entropy and Naive Bayes models. LSTM model achieved better results than the other two.

Ruchansky et al. \cite{NLP-ruchansky2017csi} proposed a model named CSI which classify an article based on temporal pattern and behavior of a user activity on the given article. Recurrence Neural Network (RNN) was used to build the CSI model. Results presented show CSI achieving a higher accuracy than the other techniques compared in the paper.

Hamid et al. \cite{NLP-karimi2018multi} proposed a technique that use Convolutional Neural Network (CNN) to extract features from a text and then apply LSTM on these features to capture the temporal dependency in the entire text. Accuracy achieved during the experiments carried out in the paper is low.

Borges et al. \cite{NLP-borges2019combining} proposed a technique based on bi-directional RNN for stance detection. First, they combine representations inferred from three different inputs from the text, including the headline, the first two sentences of the news article, and the entire document. Then, a final layer processes these results and predicts the stance of the news article.

\subsection{Discussion}

Table \ref{table:synopsis} gives a synopsis of the current works discussed in Section \ref{sec:related-work} that use machine learning techniques for automatically detecting fake news.

\begin{table*}[ht]
	\caption{A synopsis of the current works discussed in Section \ref{sec:related-work} that use machine learning (ML) techniques for automatic detection of fake news.}
	\setlength{\tabcolsep}{3pt}
	\renewcommand{\arraystretch}{1.4}
	\centering
	\begin{tabular}{ | l | m{6.5cm} | c | >{\centering}m{3.1cm} | c | } \hline
		\textbf{Research work} & \textbf{Features extraction} &  \textbf{ML methods used} & \textbf{Dataset} & \textbf{Accuracy} \tabularnewline \hline \hline
		
		Kai et al. \cite{shu2019beyond}	&	Tri-relationship among publishers, news pieces and users. Each user is assigned a credibility score based on the user's online behavior.  &    Linear classifier     & FakeNewsNet \cite{DATASET_shu2018fakenewsnet, DATASET_shu2017fake} & 87.8\% \tabularnewline \hline
		
		Yang et al. \cite{yang2018early} &  News propagation path through social media and capturing the global and local variations of user characteristics. &    Neural networks    & Weibo \cite{DATASET_WEIBO_ma2016detecting} and Twitter \cite{DATASET_TWITTER_ma2017detect} & 92\% \tabularnewline \hline

		Volkova et al. \cite{volkova2017separating} & Linguistic and network features.  &    Neural networks     & Twitter corpus \cite{DATASET_volkova2017separating} & 95\% \tabularnewline \hline

		Siering et al. \cite{siering2016detecting} & Linguistic and content based features related to different types of communication (dynamic and static). & SVM classifier & Kickstarter \cite{DATASET_kickstarter} & 80\% \tabularnewline \hline

		Afroz et al. \cite{afroz2012detecting} & Stylometry (lexical, syntactic, grammar, and author's attributes etc) to detect deception in online writing. & SVM classifier & Brennan-Greenstadt \cite{DATASET_Brennan-Greenstadt}, Hemingway-Faulkner \cite{DATASET_Hemingway-Faulkner} and Thomas-Amina \cite{DATASET_Thomas-Amina} & 96.6\% \tabularnewline \hline
		
		Hannah et al. \cite{NLP-rashkin2017truth} & Linguistic characteristics of untrustworthy text. & NLP and Neural networks & PolitiFact \cite{DATASET_politifact} & 56\% (F-score) \tabularnewline \hline
		
		Ruchansky et al. \cite{NLP-ruchansky2017csi} & Temporal pattern and behavior of a user activity on the given article. & NLP and Neural networks & Weibo \cite{DATASET_WEIBO_ma2016detecting} & 95.3\% \tabularnewline \hline
		
		Hamid et al. \cite{NLP-karimi2018multi} & Local (similar to n-grams) and global features (capturing temporal dependencies in the entire text).     &  NLP and Neural networks       & LIAR \cite{wang-2017-liar-dataset} & 38.8\% \tabularnewline \hline

		Borges et al. \cite{NLP-borges2019combining} & Combine representations inferred from the headline, first two sentences of news article, and entire document.     & NLP and Neural networks        & NLIs \cite{DATASET_NLI_bowman2015large, DATASET_NLI_williams2017broad} and FNC-1 \cite{DATASET_FNC-1} & 83.38\% \tabularnewline \hline

	\end{tabular}
	\label{table:synopsis}
\end{table*}

Out of the nine works examined, four combine NLP with deep learning, two use only deep learning, and three use other classifiers for automatic detection of fake news. Accuracy of detection varies from 38.8\% -- 96.6\%. The most important factor among these research works that effects their accuracy is the extraction and selection of features. Other factors are the type of dataset used for training and the ML method used to build the model. Less than half (four) of these research works achieve an accuracy $>$ 90\%. This shows that there is a need to further this research for improving the automatic detection of fake news.

\section{Conclusion}\label{sec:conclusion}

The phenomenon \emph{news in my pockets} is very useful in making people aware and knowledgeable of the events and their surroundings, but also presents new challenges of \emph{sieving fake news from genuine} out of this large amount of data in our pockets. In this paper, we have made an effort to present a synopsis that first explains what are fake news with examples, and then discuss some of the current machine learning techniques, such as NLP and deep learning, for automatic detection of fake news. Based on this synopsis, we recommend that there is a potential of using NLP and deep learning to improve automatic detection of fake news, but with the right set of data and features. In future work, we will explore the use of NLP with other machine learning techniques, such as neural networks and other classifiers, to automate and improve fake news detection.






%



\bibliographystyle{IEEEtran}
\bibliography{FakeNews}

\end{document}